\documentclass[twoside,twocolumn,tightenlines, showpacs,aps,prb]{revtex4}
\usepackage{graphicx}
\newcommand{\be}{\begin{equation}}
\newcommand{\ee}{\end{equation}}
\begin{document}

\title{Universal Spin Hall Conductance Fluctuations in Chaotic Dirac Quantum Dots}

\author{T. C. Vasconcelos $^1$, J. G. G. S. Ramos$^2$, A. L. R. Barbosa$^1$ }

\affiliation{$^1$Departamento de F\'{\i}sica,Universidade Federal Rural de Pernambuco, 52171-900 Recife, Pernambuco, Brazil\\
$^2$ Departamento de F\'isica, Universidade Federal da Para\'iba, 58051-970 Joa\~ao Pessoa, Para\'iba, Brazil}

\date{\today}

\begin{abstract}
We present complete analytical and numerical results that demonstrate the anomalous universal fluctuations of the spin-Hall conductance in chiral materials such as graphene and topological insulators. We investigated both the corresponding fluctuations, the Universal Fractionated and the Universal Quantized, and also the open channels orbital number crossover between the two regimes. In particular, we show that the Wigner-Dyson symmetries does not properly describe such conductances and the preponderant role of the chiral classes on the Dirac quantum dots. The results are analytical and solve outstanding issues.
\end{abstract}

\pacs{73.23.-b,73.21.La,05.45.Mt}

\maketitle
{\it Introduction} - The electronic transport through diffusive and ballistic mesoscopic devices has long been the subject of many theoretical [\onlinecite{Beenakker, Mello}] and experimental investigations [\onlinecite{Heinzel}]. The mesoscopic transport engenders some peculiar physical features [\onlinecite{Beenakker, Mello}] with special highlight to the universal conductance fluctuations (UCF) [\onlinecite{Altshuler,Stone,Ramos,BarbosaRamos}]. The UCFs occur owing to quantum interferences and chaotic scattering processes, which give rise to the sample-to-sample fluctuations of charge conductances. It was found that that in metals and semiconductors the electronic transport description is accommodated in Wigner-Dyson universal classes in the framework of Random Matrix Theory (RMT) [\onlinecite{Beenakker,Mehta}]. The universal classes are characterized by the presence or absence of two fundamental symmetries of the nature: time-reversal (TRS) and spin­rotation symmetries (SRS) [\onlinecite{Mehta}].

However, the control [\onlinecite{Wehling}] of the novel Dirac materials (as graphene and topological insulators) introduces new fundamental symmetries, as chiral/sub­lattice/mirror (CHS) and particle\-hole symmetries, which became protagonists of a myriad of interesting quantum effects [\onlinecite{jaquodbuttiker}]. Accordingly, novel RMT ensembles emerge, which are known as Chiral [\onlinecite{Verbaarschot}] and Atland-Zimbauer [\onlinecite{Zirnbauer}] universal classes. The Chiral universal classes can be applied to bipartite systems as hexagonal and square lattices whose main examples are graphene structures and topological  insulators, respectively. There are three Chiral classes: chiral circular  orthogonal  ensemble (chCOE), characterized by the presence of CHS, TRS and SRS ($\beta=1$), chiral circular unitary ensemble (CUE), which preserve CHS and has the TRS broken by external magnetic field ($\beta=2$), and chiral circular symplectic ensemble (chCSE), which is characterized by the presence of CHS, TRS and has the SRS broken by a strong spin-orbit interaction (SOI) ($\beta=4$). Moreover, Atland-Zimbauer universal classes can be applied to electronic devices in contact with a superconductor.

With the generation and control of pure spin current through mesoscopic devices, investigations on its universal fluctuations became very compelling [\onlinecite{Sarma}]. Motived by Ref.[\onlinecite{Souma}], which found the survival of the spin-Hall effect (SHE) in disordered 2D mesoscopic devices, Ren {\it et al.} [\onlinecite{Guo2006}] show numerically, using tight-binding Hamiltonians models, that diffusive mesoscopic samples with strong SOI exhibit Universal Fractionated Spin-Hall Conductance Fluctuations (UFSCF). The authors find an universal amplitude given by $\text{rms}[G_{sH}^f]\approx 0.18\; e/4\pi$. In order to give a explanation in the framework of RMT, Bardarson {\it et al.} [\onlinecite{Jacquod}] used a Landauer-B\"{u}ttiker approach and Wigner-Dyson universal class with strong SOI to obtain a analytical expression for UFSCF of a chaotic quantum dot, confirming the result of Ref.[\onlinecite{Guo2006}].

Furthermore, the 2D Dirac material exhibits the famous {\it quantized} spin-Hall effect, which mean that spin-Hall conduction $G_{sH}^q$ takes only integers multiples of $e/4\pi$ [\onlinecite{Kane,Abanin}]. Motivated by this novel feature, Qiao {\it et al.} [\onlinecite{Guo2008}] show numerically, using three tight-binding Hamiltonians models, that the diffusive 2D Dirac samples with and without strong SOI exhibit Universal Quantized Spin-Hall Conductance Fluctuations (UQSCF) with an amplitude given by $\text{rms}[G_{sH}^q]=0.285\pm 0.005\; e/4\pi$, which does not follow the conventional value obtained in Refs.[\onlinecite{Guo2006,Jacquod}]. In recent work, Choe and Chang [\onlinecite{Chang}] study electronic transport in $2D$ Dirac device with strong SOI and CHS numerically. Nevertheless, their studies were not enough to reach a definitive understanding of the behavior in terms of a UQSCF obtained in Ref.[\onlinecite{Guo2008}]. Thus an understanding of UFSC remains open to date.

In this work, we investigate analytically the UQSCF in a chaotic quantum dot with CHS, also known as chaotic Dirac quantum dot [\onlinecite{Ponomarenko,Barros}], at low temperature. We assume a preserved TRS, SRS broken by a strong SOI (chCSE), and a mean dwell time of electrons in the quantum dot that is larger than the SOI time, $\tau_{dwell}\gg\tau_{so}$. We identified two regimes: the first one, when the CHS is broken, gives rise to the UFSCF and exhibits an amplitude $\text{rms}[G_{sH}^f]\approx \sqrt{2}\times 0.18\; e/4\pi$; The second one, when CHS is preserved, gives rise to the UQSCF, which assumes an amplitude $\text{rms}[G_{sH}^q]\approx 0.283\; e/4\pi$. The last in agreement with the Ref.[\onlinecite{Guo2008}].

\begin{figure}
\centering
\includegraphics[width=7cm]{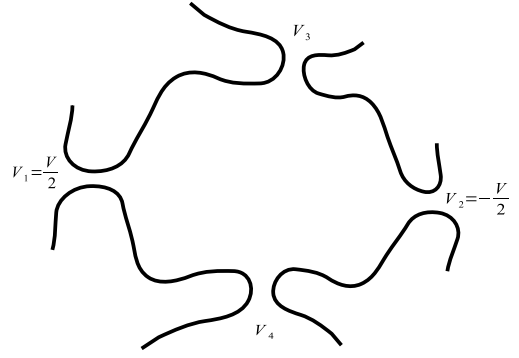}
\caption{A chaotic Dirac quantum dot connected through four leads to electron reservoirs with spin resolution. The $1$ and $2$ leads have specific potentials, $V/2$ and $-V/2$, while the $3$ and $4$ leads have potentials adjusted in a way that prevents the charge current flowing. Therefore, a charge current between the $1$ and $2$ leads induce a spin current between the $3$ and $4$ leads.}
\label{Imagem1}
\end{figure}

{\it Theorical Framework} - We consider a multi-terminal $2D$ device with CHS symmetry where the electrons flow under the influence of a strong SOI at low temperature. The $2D$ device is connected by ideal point contacts to four independent electronic reservoirs as depicted in Fig.(\ref{Imagem1}). We use the Landauer-B\"uttiker approach to write the $\alpha$-direction spin-resolved current through the terminals as [\onlinecite{Buttiker}]
\begin{equation}
I_{i}^\alpha=\frac{e^2}{h}\sum_{j,\alpha'}\tau_{i,j}^{\alpha\alpha'}\left(V_{i}-V_{j}\right).\label{Ii}
\end{equation}

The spin-dependent transmission coefficients can be obtained through $\tau_{i,j}^{\alpha\alpha'}=\sum_{m\in i,n\in j} |\mathcal{S}_{m,\alpha;n,\alpha'}|^2$, where $\alpha$ and $\alpha'$ are the spin projections in the $\alpha= x, y,$ or $z$ direction and $\mathcal{S}$ is the scattering matrix of order $2 \bar{ N}_T \times 2 \bar{N}_T$. The $\mathcal{S}$-matrix describes the transport of electrons through the chaotic Dirac quantum dot. The total number of open orbital scattering channels is $\bar{N}_T= 2 N_T = \sum_{i=1}^4 2 N_i$, where $2N_i$ is the number of open channels in $i$th lead point contact and the factor $2$ came from two sublattices of the Dirac Materials [\onlinecite{beenakkergrafeno}].

An applied bias voltage $V$ between longitudinal electrodes $1$ and $2$ gives rise to a longitudinal electronic current $I$ and, due to the {\it quantized} spin Hall effect, to spin currents $I^{\alpha}_i= I^{1}_i - I^{1}_i\neq 0 $ at the transversal contacts $3$ and $4$ with $\alpha=x,y,z$ [\onlinecite{RamosBarbosa}], as depicted in Fig.\ref{Imagem1}. Moreover, we consider the absence of net electric charge current at the transverse $3$ and $4$ leads, i.e, $I^0_i = I^{1}_i + I^{1}_i=0$. Therefore, the charge conservation implies $I = I^0_1=-I^0_2$. Using those constraints in Eq.(\ref{Ii}), it was shown in Ref.[\onlinecite{Jacquod}] that the transversal spin currents can be written as
\begin{equation}
J_{i}^\alpha=\frac{1}{2}(\tau_{i2}^{\alpha}-\tau_{i1}^{\alpha})-\sum_{j=3,4}\tau_{ij}^{\alpha}\;\bar{V_j}, \label{Ji}
\end{equation}
for which we introduce the dimensionless currents $J = h/e^2(I/V )$ and the effective transverse voltages $\bar{V}_{i} = V_i/V$, given by
\begin{eqnarray}
\bar{V}_{i}&=&\frac{1}{2}\frac{\tau_{ij}^0(\tau_{j2}^0-\tau_{j1}^0)+(\tau_{i2}^0-\tau_{i1}^0)(4N_j-\tau_{jj}^0)}{\tau_{43}^0\tau_{34}^0-(4N_3-\tau_{33}^0)(4N_4-\tau_{44}^0)} \label{V}
\end{eqnarray}
where here $i,j = 3,4$ with $i \neq j$.

{\it Random Matrix Theory} - Our calculation consists in the obtention of the average and the fluctuation amplitude of transversal spin currents, Eq.(\ref{Ji}), for a chaotic Dirac quantum dot, within the framework of RMT. Therefore, the spin-dependent transmission coefficient can be written in an appropriate way thought the following relation
\begin{equation}
\tau_{ij}^\alpha =\textbf{Tr}[\mathcal{P}^\alpha_{i}\mathcal{S}\mathcal{P}^0_{j}\mathcal{S}^{\dagger}], \qquad \alpha = 0,x,y,z \label{tauij}
\end{equation}
where the scattering matrix $\mathcal{S}$ is a member of the  chCSE ensemble, which means the system preserve the TRS (absence of magnetic field) and the SRS is broken by strong SOI \cite{Verbaarschot}. The matrix $\mathcal{P}^\alpha_i = \mathcal{P}_i \otimes \sigma^\alpha$ represents a projector operator over $i$th terminal. Its dimension is $2 \bar{ N}_T \times 2 \bar{N}_T$ and its entries are $\left(\mathcal{P}^\alpha_i\right)_{m\mu,n\gamma}=\delta_{mn} \sigma^\alpha_{\mu\gamma},$ while $\sum_{j=1}^{i-1}2N_j< m <\sum_{j=1}^{i}2N_j$ and $0$ for otherwise\cite{Jacquod}. The $\sigma^0$ and $\sigma^\alpha$ are the identity matrix and Pauli matrices, respectively.

The scattering matrix of Eq.(\ref{tauij}) has the additional CHS symmetry, implying it satisfies the following commutation relation  \cite{jaquodbuttiker}
\begin{eqnarray}
\mathcal{S}=\Sigma_{z}\mathcal{S}^{\dagger} \Sigma_{z}, \quad
\Sigma_{z}\equiv
 \left[\begin{array}{cc}
\textbf{1}_{\bar{N}_T}  & 0\\
0 & -\textbf{1}_{\bar{N}_T}
 \end{array}
 \right] \label{S}
 \end{eqnarray}
at the Dirac point (null Fermi energy). To obtain the average of Eq.(\ref{tauij}), it is convenient to decompose $\mathcal{S}$ as a function of the $\mathcal{U}$-matrix, which is a symplectic matrix of order $2 \bar{ N}_T \times 2 \bar{N}_T$, as $\mathcal{S}=\Sigma_z\mathcal{U}^\dagger\Sigma_z\mathcal{U}$, as can be seen in Ref.[\onlinecite{Nish}]. Hence, the Eq.(\ref{tauij}) can be rewritten as
\begin{equation}
\tau_{ij}^\alpha =\textbf{Tr}[\mathcal{P}^\alpha_i\mathcal{U}^\dagger\Sigma_z\mathcal{U}\mathcal{P}^0_j\mathcal{U}^\dagger\Sigma_z\mathcal{U}] \qquad \alpha = 0,x,y,z.\label{tauijU}
\end{equation}

Using the method of Ref.[\onlinecite{Barros}], developed for the diagrammatic integration over chaotic Dirac quantum dots, we calculate the average and covariance of spin-dependent transmission coefficient of Eq.(\ref{tauijU}). Firstly, for the average of Eq.(\ref{tauijU}), we obtain
\begin{equation}
\langle\tau_{ij}^\alpha\rangle =4\delta_{\alpha0}\frac{N_T(4N_iN_j-\delta_{ij}N_i)+\delta_{ij}N_i}{(2N_T-1)(N_T+1)}.\label{Mtauij}
\end{equation}
The Eq.(\ref{Mtauij}) is quite distinct from the result of Ref.[\onlinecite{Jacquod}], which studied the UFSCF of a Schr\"odinger (Wigner-Dyson) chaotic quantum dot. Secondly, following the same method of Ref.[\onlinecite{Barros}], the covariance of Eq.(\ref{tauijU}) reads $$\text{covar}[\tau_{ij}^{\alpha},\tau_{kl}^{\beta}]= \left\langle\left( \tau_{ij}^{\alpha}-\left\langle\tau_{ij}^{\alpha}\right\rangle\right) \left( \tau_{kl}^{\beta}-\left\langle\tau_{kl}^{\beta}\right\rangle\right) \right\rangle$$ and prompted us to find $11025$ diagrams, from which $6300$ are non-null. Therefore, the overall result for the covariance of spin-dependent transmission coefficient involves $6300$ terms and the algebraic final expression is cumbersome. Nevertheless, for the relevant configuration wherein $\alpha=\beta \neq 0$ and $i=k$, the general expression simplifies to
\begin{eqnarray}
\frac{\text{covar}[\tau_{ij}^{\alpha},\tau_{il}^{\alpha}]}{D}=\left\{ \matrix{
(N_T-N_i)(4N_iN_T-2N_T-3), i=j=l \cr
-4N_jN_lN_T, i\neq{j}\neq{l} \cr
N_j(4N_T^2-4N_jN_T-2N_T-3), i\neq{j}=l \cr
-N_l(4N_iN_T-2N_T-3), i=j\neq{l} \cr
-N_j(4N_iN_T-2N_T-3), i=l\neq{j} \cr}\right.
\label{mean}
\end{eqnarray}
where $D=\frac{128N_iN_T}{(4N_T+3)(16N_T^2-1)(2N_T-3)(2N_T-1)}$.

{\it Universal Spin Hall Conductance Fluctuations} - We begin using the Eq.(\ref{Mtauij}) to obtain the ensemble average of the transverse spin currents of Eq.(\ref{Ji}). The result is
\begin{equation}
\langle{J_{i}^\alpha}\rangle=\frac{1}{2}
(\langle\tau_{i2}^\alpha\rangle-\langle\tau_{i1}^\alpha\rangle)-\sum_{j=3,4}\langle\tau_{ij}^{\alpha}\rangle\langle\bar{V_j}\rangle =0 \label{Jim}
\end{equation}
as expected even for $i=3,4$ and $\alpha = x,y,z$. The ensemble average of the effective transverse voltages, Eq.(\ref{V}), is obtained with the previous spin-dependent transmission coefficient general result for the covariance. We obtain
\begin{equation}
\langle\bar{V}_{3,4}\rangle=\frac{1}{2}\frac{N_1-N_2}{N_1+N_2}.\label{V34}
\end{equation}
The same expression was obtained in Ref.[\onlinecite{Jacquod}], indicating that the effective transverse voltage is CHS independent.

\begin{figure}
\centering
\includegraphics[width=8cm]{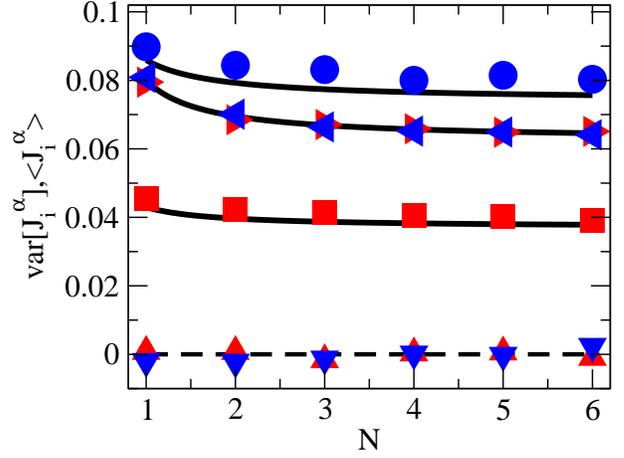}
\caption{Average (\ref{Jim}) and variance (\ref{main}) of transverse spin current $J_i^\alpha$ for $i=3,4$ and $\alpha=x,y,z$ as a function of open channels $N$. The analytical results are represented by solid lines, while the numeric simulations, obtained by RMT, are represented by the symbols. Symmetric terminals ($N_1=N_2=N_3=N_4=N$): (triangle left) $\text{var}[J^\alpha_{3}]$, and (triangle right) $\text{var}[J^\alpha_{4}]$. Asymmetric terminals ($N_1=N_3=2N_2=2N_4=N$): (triangle up) $\langle{J_{3}^\alpha}\rangle$, (triangle down) $\langle{J_{4}^\alpha}\rangle$, (square) $\text{var}[J^\alpha_{3}]$ and (circle) $\text{var}[J^\alpha_{4}]$}
\label{Imagem2}
\end{figure}

Although the transversal spin currents ensemble average vanishes, the Eq.(\ref{V34}) establishes an effective non-null transverse voltage. Accordingly, the transversal spin current $J_{i}^\alpha$ assumes finite values. The amplitude of the fluctuations of $J_{i}^\alpha$ is given by\cite{Jacquod}
\begin{eqnarray}
\text{var}[J^\alpha_i]&=&\frac{1}{4}\sum_{j=1,2}\text{var}[\tau_{ij}^\alpha]-\frac{1}{2}\sum_{j=1,2}\text{covar}[\tau_{i1}^\alpha,\tau_{i2}^\alpha]
\nonumber\\&+&\sum_{j=3,4}(\text{var}[\tau_{ij}^\alpha]\langle\bar{V^2_j}\rangle
+\text{covar}[(\tau_{i1}^\alpha-\tau_{i2}^\alpha),\tau_{ij}^\alpha]\langle\bar{V_j}\rangle)
\nonumber\\&+&2\text{covar}[\tau_{i3}^\alpha,\tau_{i4}^\alpha]\langle\bar{V_3}\rangle\langle\bar{V_4}\rangle.\label{varJ}
\end{eqnarray}
Substituting the Eqs.(\ref{mean}) and (\ref{V34}) in Eq.(\ref{varJ}), we obtain the following expression
\begin{equation}
\text{var}[J^\alpha_{i}]=\frac{128N_iN_1N_2N_T(4N^2_T-2N_T-3)}{(4N_T+3)(16N^2_T-1)(2N_T-3)(2N_T-1)(N_1+N_2)},\label{main}
\end{equation}
which is the main result of our work. The Eq.(\ref{main}) is quite distinct from the main result of Ref.[\onlinecite{Jacquod}], and reveals the full difference between the spintronics of a chaotic Dirac (CHS) quantum dot and a chaotic Schr\"odinger (Wigner-Dyson) quantum dot. In the Fig.(\ref{Imagem2}), we plot the average (\ref{Jim}) and variance (\ref{main}) of transverse spin current $J_i^\alpha$ for $i=3,4$ and $\alpha=x,y,z$ as a function of both symmetric open leads, $N_1=N_2=N_3=N_4=N$, and asymmetric open leads, $N_1=N_3=2N_2=2N_4=N$.

Using the Eq.(\ref{main}), we can analyse two relevant regimes of the chaotic Dirac quantum dots: The first one for the broken CHS, Eq.(\ref{S}), and the second one when it is preserved. In accordance with the Refs.[\onlinecite{Gnutzmann,Richter}], the CHS is relevant for the systems at zero Fermi energy and/or if there are few open channels in the leads. However, if the Fermi energies are away from zero and/or if there are many open channels in the terminals, the Wigner-Dyson universality classes and the Chiral universality classes lead presumably to the same results.

We first investigate the setup with broken CHS employed whenever the system has a large number of open channels or high Fermi energy. For this system, we fix symmetric terminals, $N_1=N_3=N_2=N_4=N$, from the general result with a large number of open channels $N\gg 1$ in Eq.(\ref{main}), and we obtain $\text{var}[J^\alpha_{i}]=2 \times 1/32$. Therefore, the spin current fluctuates universally with amplitude $\text{rms}[I_{3}^z]=0.25\; e^2V/h$, which can be used to write the universal fluctuations of transversal spin conductance as
\begin{equation}
\text{rms}[G_{sH}^f]\approx\sqrt{2}\times0.18\;\frac{e}{4\pi}.\label{gshsc}
\end{equation}
In this regime, the universal conductance fluctuations has amplitude $\sqrt{2}$ times higher than the result obtained in Refs.[\onlinecite{Guo2006,Jacquod}] for the Wigner-Dyson universality classes. Furthermore, the result is in agreement with Ref.[\onlinecite{Guo2008}] when the Fermi energies are $|E|>1$ and with the studies concerning to universal conductance fluctuations in two-dimensional topological insulators with strong SOI of Ref.[\onlinecite{Chang}].

\begin{figure}
\centering
\includegraphics[width=7cm]{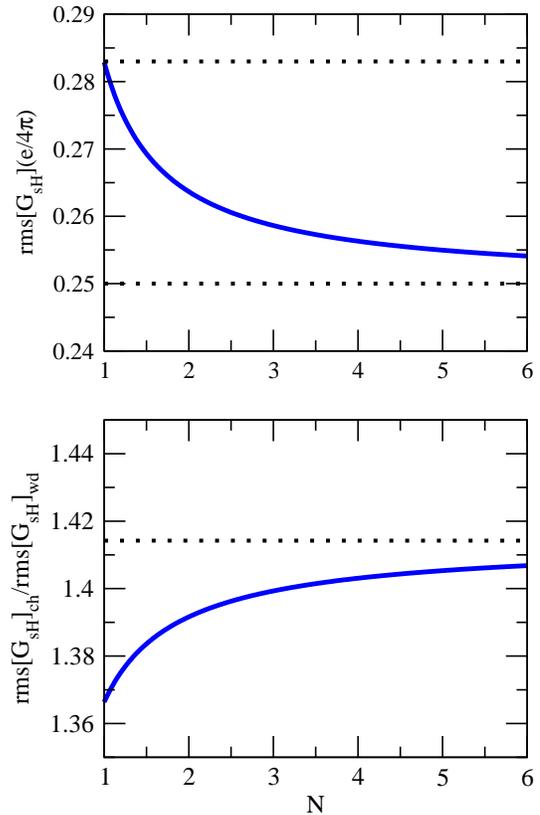}
\caption{(Up) The universal fluctuations of transversal spin conductance plotted as a open channels crossover from Eq.(\ref{gshh}) to Eq.(\ref{gshsc}). (Down) The ratio between universal fluctuations of transversal spin conductances of Chiral universal classes and Wigner-Dyson universal classes one $(\text{rms}[G_{sH}]_{ch}/\text{rms}[G_{sH}]_{wd}$); the $\sqrt{2}$ limit as a function of $N$.}
\label{Imagem3}
\end{figure}

However, without the CHS symmetry (Wigner-Dyson ensembles), the universal fluctuations of transversal spin conductance of a chaotic Dirac quantum dot does not describe the result obtained in Ref.[\onlinecite{Guo2008}], i.e $\text{rms}[G_{sH}^q]=(0.285\pm 0.005)\; e/4\pi$, complied only if the number of channels is very small (quantized). For this reason, we fix from our  general analytical result a small number of channels in order to preserve the CHS. Using symmetric configuration with $N_1=N_3=N_2=N_4=1$, known as high quantum regime, in Eq.(\ref{main}), we obtain
$\text{rms}[I_{3}^z]\approx0.283\; e^2V/h$. The result can be rewritten in terms of the universal fluctuations of transversal spin conductance as
\begin{equation}
\text{rms}[G_{sH}^q]\approx0.283\;\frac{e}{4\pi} \label{gshh}
\end{equation}
which is agreement with Ref.[\onlinecite{Guo2008}] if the Fermi energies are $|E|<1$. In the Fig.(\ref{Imagem3})-up, we plot the universal fluctuations of transversal spin conductance as a function of $N$, which shows the break of CHS as a open channels crossover from Eq.(\ref{gshh}) to Eq.(\ref{gshsc}).

Importantly, the result of Eq.(\ref{gshh}) for the the variance of transversal spin current is incompatible with the one of Ref.[\onlinecite{Jacquod}], obtained for the Wigner-Dyson universal classes. In high quantum regime, the result of Ref.[\onlinecite{Jacquod}] multiplied for $\sqrt{2}$ yields $\sqrt{2}\times\text{rms}[G_{sH}]_{wd}\approx0.292\;e/4\pi$, therefore outside the error bar of the tight-biding simulation of Ref.[\onlinecite{Guo2008}] valid for graphene and another chiral systems. In the Fig.(\ref{Imagem3})-down, we depicted the ratio between universal fluctuations of transversal spin conductances of the Chiral universal classes and the Wigner-Dyson ones. Notice the ratio $(\text{rms}[G_{sH}]_{ch}/\text{rms}[G_{sH}]_{wd}$) tends to $\sqrt{2}$ as a function of $N$ from the general analytical result.

{\it Numeric Simulation} - In order to confirm the analytical results, Eqs.(\ref{Jim}) and (\ref{main}), we use a numerical simulations from the RMT \cite{Verbaarschot}.

The massless Dirac Hamiltonian satisfy the following anti-commutation relation\cite{jaquodbuttiker,Verbaarschot}
\begin{eqnarray}
\mathcal{H}= -\lambda_{z}\mathcal{H}\lambda_{z}, \quad
\lambda_{z}=
 \left[
\begin{array}{cc}
\textbf{1}_{2M}  & 0\\
0 & -\textbf{1}_{2M}
 \end{array}
 \right].\label{H}
\end{eqnarray}
The $\mathcal{H}$-matrix has dimension $4 M\times 4 M$. The anti-commutation relation, Eq.(\ref{H}) implies a Hamiltonian member that can be written as
\begin{eqnarray}
\mathcal{H}=\left(
\begin{array}{cc}
\textbf{0} & \mathcal{T} \\
\mathcal{T}^{\dagger} & \textbf{0}
\end{array}
\right).
\label{H1}
\end{eqnarray}
The quaternionic $\mathcal{T}$-matrix block of the $\mathcal{H}$-matrix has dimension $2M\times 2M$. The RMT establishes that the entries of $\mathcal{T}$-matrix have Gaussian distribution
\begin{eqnarray}
P( \mathcal{T} )\propto \exp\left\lbrace -\frac{2 M}{\lambda^{2}}Tr( \mathcal{T} ^{\dagger} \mathcal{T} )\right\rbrace,
\end{eqnarray}
where $\lambda=4M\Delta/\pi$ is the variance related to the electronic single-particle level spacing, $\Delta$. The Hamiltonian model for the scattering matrix can be written as\cite{Weidenmuller}
\begin{eqnarray}
\mathcal{S}=\textbf{1}-2\pi i\mathcal{W}^{\dagger}(\epsilon-\mathcal{H} +i\pi\mathcal{W}\mathcal{W}^{\dagger})^{-1}\mathcal{W},\label{MW}
\end{eqnarray}
which satify the Eq. (\ref{S}).
The $\mathcal{W} = \textbf{(}\mathcal{W}_{1},\dots, \mathcal{W}_{4}\textbf{)} $ matrix is a $4M \times 2\bar{N}_T$ deterministic matrix, describing the coupling of the resonances states of the chaotic Dirac quantum dot with the propagating modes in the four terminals. This deterministic matrix satisfies non-direct process, i.e., the orthogonality condition $ \mathcal{W}_{p}^{\dagger}\mathcal{W}_{q}=\frac{1}{\pi}\delta_{p,q}$ holds. Accordingly, we consider the relation $\lambda_z \mathcal{W}\Sigma_z=\mathcal{W}$, indicating the scattering matrix is symmetric (\ref{S}). We consider the system on the Dirac point ($\epsilon=0$), and, to ensure the chaotic regime and consequently the universality of the observable, the number of resonances inside quantum dot is large ($M\gg N_T$).

The numerical simulations produce Fig.(\ref{Imagem2}), which shows symbols obtained through $25\times 10^3$ realizations compared with the analytical results, the Eqs.(\ref{Jim}) and (\ref{main}). We use the $\mathcal{T} $ matrices, with dimension $800\times 800$ ($M=400$), and the corresponding $\mathcal{H}$ matrices with dimension $1600 \times 1600$ ($1600$ resonances).

{\it Conclusions} - To summarize, we present a complete analytical study of UQSCF of chaotic Dirac quantum dot in the framework of RMT for the Chiral universal symmetries in the absence of magnetic field. We show that the effective transverse voltage is CHS independent, Eq.(\ref{V34}). Moreover, in the regime of broken CHS the UFSCF exhibit a value $\text{rms}[G_{sH}^f]\approx \sqrt{2}\times 0.18\; e/4\pi$. However, the system with preserved CHS exhibits a UQSCF with amplitude $\text{rms}[G_{sH}^q]\approx 0.283\; e/4\pi$, which is in agreement with the numerical simulation of Ref.[\onlinecite{Guo2008}]. We can conclude that quantized spin-Hall effect in the absence of magnetic field is described by Chiral universal classes in the framework of RMT. Perspectives of our work include the study of the Universal Classes of the UQSCF in the presence of magnetic field preserving the particle-hole symmetry.

This work was partially supported by CNPq, CAPES and FACEPE (Brazilian Agencies).

\end{document}